\documentclass[12pt]{article}
\usepackage{amsfonts}
\usepackage{amssymb}

\def\hybrid{\topmargin -20pt    \oddsidemargin 0pt
        \headheight 0pt \headsep 0pt
        \textwidth 6.25in       
        \textheight 9.25in       
        \marginparwidth .875in
        \parskip 5pt plus 1pt   \jot = 1.5ex}

\hybrid

\def\baselinestretch{1.2}

\catcode`\@=11

\def\marginnote#1{}
\newcount\hour
\newcount\minute
\newtoks\amorpm
\hour=\time\divide\hour by60
\minute=\time{\multiply\hour by60 \global\advance\minute by-\hour}
\edef\standardtime{{\ifnum\hour<12 \global\amorpm={am}%
        \else\global\amorpm={pm}\advance\hour by-12 \fi
        \ifnum\hour=0 \hour=12 \fi
        \number\hour:\ifnum\minute<10 0\fi\number\minute\the\amorpm}}
\edef\militarytime{\number\hour:\ifnum\minute<10 0\fi\number\minute}

\def\draftlabel#1{{\@bsphack\if@filesw {\let\thepage\relax
   \xdef\@gtempa{\write\@auxout{\string
      \newlabel{#1}{{\@currentlabel}{\thepage}}}}}\@gtempa
   \if@nobreak \ifvmode\nobreak\fi\fi\fi\@esphack}
        \gdef\@eqnlabel{#1}}
\def\@eqnlabel{}
\def\@vacuum{}
\def\draftmarginnote#1{\marginpar{\raggedright\scriptsize\tt#1}}

\def\draft{\oddsidemargin -.5truein
        \def\@oddfoot{\sl preliminary draft \hfil
        \rm\thepage\hfil\sl\today\quad\militarytime}
        \let\@evenfoot\@oddfoot \overfullrule 3pt
        \let\label=\draftlabel
        \let\marginnote=\draftmarginnote
   \def\@eqnnum{(\theequation)\rlap{\kern\marginparsep\tt\@eqnlabel}%
\global\let\@eqnlabel\@vacuum}  }

\def\preprint{\twocolumn\sloppy\flushbottom\parindent 2em
        \leftmargini 2em\leftmarginv .5em\leftmarginvi .5em
        \oddsidemargin -.5in    \evensidemargin -.5in
        \columnsep .4in \footheight 0pt
        \textwidth 10.in        \topmargin  -.4in
        \headheight 12pt \topskip .4in
        \textheight 6.9in \footskip 0pt
        \def\@oddhead{\thepage\hfil\addtocounter{page}{1}\thepage}
        \let\@evenhead\@oddhead \def\@oddfoot{} \def\@evenfoot{} }

\def\numberbysection{\@addtoreset{equation}{section}
        \def\theequation{\thesection.\arabic{equation}}}

\def\underline#1{\relax\ifmmode\@@underline#1\else
        $\@@underline{\hbox{#1}}$\relax\fi}

\def\titlepage{\@restonecolfalse\if@twocolumn\@restonecoltrue\onecolumn
     \else \newpage \fi \thispagestyle{empty}\c@page\z@
        \def\thefootnote{\fnsymbol{footnote}} }

\def\endtitlepage{\if@restonecol\twocolumn \else \newpage \fi
        \def\thefootnote{\arabic{footnote}}
        \setcounter{footnote}{0}}  

\catcode`@=12
\relax

\def\figcap{\section*{Figure Captions\markboth
        {FIGURECAPTIONS}{FIGURECAPTIONS}}\list
        {Figure \arabic{enumi}:\hfill}{\settowidth\labelwidth{Figure
999:}
        \leftmargin\labelwidth
        \advance\leftmargin\labelsep\usecounter{enumi}}}
 \relax
\def\tablecap{\section*{Table Captions\markboth
        {TABLECAPTIONS}{TABLECAPTIONS}}\list
        {Table \arabic{enumi}:\hfill}{\settowidth\labelwidth{Table
999:}
        \leftmargin\labelwidth
        \advance\leftmargin\labelsep\usecounter{enumi}}}
 \relax
\def\reflist{\section*{References\markboth
        {REFLIST}{REFLIST}}\list
        {[\arabic{enumi}]\hfill}{\settowidth\labelwidth{[999]}
        \leftmargin\labelwidth
        \advance\leftmargin\labelsep\usecounter{enumi}}}
 \relax
\makeatletter
\newcounter{pubctr}
\def\publist{\@ifnextchar[{\@publist}{\@@publist}}
\def\@publist[#1]{\list
        {[\arabic{pubctr}]\hfill}{\settowidth\labelwidth{[999]}
        \leftmargin\labelwidth
        \advance\leftmargin\labelsep
        \@nmbrlisttrue\def\@listctr{pubctr}
        \setcounter{pubctr}{#1}\addtocounter{pubctr}{-1}}}
\def\@@publist{\list
        {[\arabic{pubctr}]\hfill}{\settowidth\labelwidth{[999]}
        \leftmargin\labelwidth
        \advance\leftmargin\labelsep
        \@nmbrlisttrue\def\@listctr{pubctr}}}
 \relax
\makeatother
\newskip\humongous \humongous=0pt plus 1000pt minus 1000pt

\newif\ifdtup

\relax

\def\be{\begin{equation}}
\def\ee{\end{equation}}
\def\ba{\begin{eqnarray}}
\def\ea{\end{eqnarray}}

\def\del{\partial}

\def\a{\alpha}

\def\b{\beta}

\def\g{\gamma}
\def\G{\Gamma}
\def\d{\delta}
\def\D{\Delta}

\def\m{\mu}
\def\n{\nu}

\def\l{\lambda}

\def\cN{{\cal N}}

\def\cL{{\cal L}}
\def\cR{{\cal R}}

\def\no{\noindent}

\def\qq{\qquad}

\def\IR{\relax{\rm I\kern-.18em R}}

\def \ha {{1\over 2}}

\def \ov {\over}

\def\IR{\relax{\rm I\kern-.18em R}}
\def\inv{^{\raise.15ex\hbox{${\scriptscriptstyle -}$}\kern-.05em 1}}

\def\cL{{\cal L}}
\def\cR{{\cal R}}

\def\tr{{\rm tr}}

\begin{document}

\renewcommand{\theequation}{\arabic{equation}}

\newcommand{\beq}{\begin{equation}}
\newcommand{\eeq}[1]{\label{#1}\end{equation}}
\newcommand{\ber}{\begin{eqnarray}}
\newcommand{\eer}[1]{\label{#1}\end{eqnarray}}
\newcommand{\eqn}[1]{(\ref{#1})}
\begin{titlepage}
\begin{center}

\hfill CALT-68-2381\\
\vskip -.1 cm
\hfill CITUSC-02-012\\
\vskip -.1 cm
\hfill hep--th/0204193\\

\vskip .5in

{\large \bf Current correlators and AdS/CFT}\\
{\large\bf away from the conformal point}${}^{}$\footnote{
Contribution to the proceedings of the {\it 7th Hellenic summer school
and Workshops on High Energy Physics},
Corfu, Greece, 13 August-13 September 2001; based on an invited
lecture given by K.S.}

\vskip 0.4in

{\bf Andreas Brandhuber}${}^1$
and
{\bf Konstadinos Sfetsos}${}^2$
\vskip 0.1in
{\em ${}^1\!$Department of Physics\\
     California Institute of Technology\\
     Pasadena, CA 91125, USA\\
\footnotesize{\tt andreas@theory.caltech.edu}}\\
\vskip .2in
{\em ${}^2\!$Department of Engineering Sciences, University of Patras\\
26110 Patras, Greece\\
\footnotesize{\tt sfetsos@mail.cern.ch, des.upatras.gr}}\\

\end{center}

\vskip .3in

\centerline{\bf Abstract}

\noindent
Using the AdS/CFT correspondence we study vacua of $\mathcal{N}=4$
SYM for which
part of the gauge symmetry is broken by expectation values of scalar
fields. A specific subclass of such vacua can be analyzed with gauged
supergravity and the corresponding domain wall solutions lift to
continuous distributions of D3-branes in type IIB string theory.
Due to the non-trivial
expectation value of the scalars, the $SO(6)$ $\mathcal{R}$-symmetry
is spontaneously broken and field theory
predicts the existence of Goldstone bosons. We explicitly show that,
in the dual supergravity description, these emerge as
massless poles in the current two-point functions, while
the bulk gauge fields which are dual to the broken currents
become massive via the Higgs mechanism.
We find agreement with field theory expectations and,
hence, provide a non-trivial test of the AdS/CFT correspondence far
away from the conformal point.

\vskip .4in
\noindent
April 2002\\
\end{titlepage}
\vfill
\eject

\def\baselinestretch{1.2}
\baselineskip 16 pt
\noindent

\section{Introduction}

Up to this day the AdS/CFT correspondence \cite{malda,gkp,witten} is the
most concrete proposal for a realization of holography.
Although this duality allows to perform numerous explicit
calculations in the supergravity limit an honest proof of
this conjecture directly from string theory is still elusive.
Therefore, the best we can do, is to take the results obtained
from supergravity as predictions and make qualitative and,
more desirably, quantitative comparisons with field theory.
Indeed, it is an interesting task in itself to study
known phenomena of quantum field theories and try to understand
how they are realized on the dual string/gravity side. Most of the
comparisons have been performed for the $\mathcal{N}=4$ SYM theory at
the conformal point.  Naturally, one may
wonder whether the AdS/CFT correspondence can be checked away from
conformality as well. This question has been answered positively in
\cite{jjcorr} where we computed 2-point functions of currents in the Coulomb
branch of the $\mathcal{N}=4$ SYM theory and
provided a non-trivial test of the AdS/CFT correspondence
in the deep infrared of the theory.
It is the purpose of this paper to review this work for the proceedings of
the ``Corfu summer school on elementary particle Physics''.
Relevant literature on the Coulomb branch of $\cN=4$ as well as other works
on current correlators within
the AdS/CFT correspondence can
be found in the reference list of our original paper \cite{jjcorr}
(particularly, in \cite{fmmr,massimo}).

In order to study theories with less supersymmetry and/or
broken conformal symmetry, which are closer to the theories
realized in Nature, deformations of the original
conjecture \cite{malda} have been studied extensively over the past years.
In this paper we are concerned with the simplest
modification of $\mathcal{N}=4$ SYM
by turning on vacuum expectation values (vevs) of
scalar fields.
The $\mathcal{N}=4$ vector multiplet contains, beside
the vector potential and four adjoint Weyl fermions, six
scalars in the adjoint representation of the gauge group $U(N)$.
These scalars transform in the ${\bf 6}$ of the $SO(6)$
$\mathcal{R}$-symmetry group and maybe represented by the $N\times N$
matrices $\Phi^i$, $i=1,2,\dots ,6$. The quartic scalar potential of
the theory $\sum_{i<j} \tr[\Phi^i,\Phi^j]^2$
has flat directions that are parametrized by
six diagonal $N \times N$ matrices
\be
X^i_{\rm vev}=\langle \Phi^i\rangle = {\rm diag}(X^i_1,X^i_2,\dots ,
X^i_N)\ , \qq
\sum_{p=1}^N X^i_p=0\ .
\label{vvee}
\ee
The corresponding Coulomb branch
is $(\mathbb{R}^6)^N/S_N$ and on points away from the origin
the conformal symmetry is broken.
The action of $\mathcal{N}=4$ SYM contains the term
$\sum_i \tr(D_\m \Phi^i)^2$
which couples the gauge fields to the scalars. When the latter acquire
vevs the gauge bosons become massive.
It is convenient to choose the standard real basis for the $SU(N)$
generators $J_{pq}=e_{pq}-1/N \d_{pq} I_{N\times N}$, where the matrix
elements of the matrices $e_{pq}$ are: $(e_{pq})_{rs}=\d_{pr} \d_{qs}$.
Hence, according to \eqn{vvee}, we give vevs to the scalars represented
by the six-dimensional vector $\vec \Phi$, as
$\vec \Phi=h_p \vec X_p$,
where $h_p=J_{pp}$ are the generators of the Cartan subalgebra of $SU(N)$.
The masses of the gauge fields arise from the term $\sum_i \tr[\Phi^i,A_\m]^2$.
After some computation we find the (mass)$^2$ matrix with elements
\cite{sfedyn,jjcorr}
\be
(M^2)_{pq}=|\vec X_p-\vec X_q |^2,\qq p,q=1,2,
\dots , N\ ,
\label{maaa}
\ee
up to a numerical factor of order 1. Hence, the masses have the
geometrical interpretation as the distances between the various
vev positions distributed in the scalar space $\mathbb{R}^6$.
Equivalently, they are given by
the masses of the strings stretched between the D3-branes located
at these points.
It is clear, that
some of these masses may be degenerate,
depending on the specific distribution of vevs.

Let us illustrate some of the features with the toy example of
a discrete distribution of vevs in an
polygon with $N$ vertices located on a circle
of radius $r_0$ in the $1$-$2$ plane \cite{sfe1}
\be
\vec X_{p} = \left( r_0 \cos \phi_p, r_0 \sin \phi_p,0,0,0,0 \right) \ ,
\quad \phi_p=2\pi p/N\ ,\quad p=1,2,\dots , N\ .
\label{lv}
\ee
In this case we find from \eqn{maaa}, that \cite{sfedyn}
\be
M_n = 2 r_0 \sin(\pi n/N)\ ,\qq n=1,2,\dots , N\ ,
\label{hwe}
\ee
which is an exact result for any $N$.
The degeneracy for the zero mode is $d_N=N-1$ and for the rest $d_n=2(N-n)$.
It is easily seen that $\sum_{n=1}^N d_n=N^2-1$.
Hence, for large $N$ there are W-bosons with masses of order $r_0$
and W-bosons with light masses of order $r_0/N$.
For more general distributions
the same result holds with $r_0$ being replaced by the average value for
the distribution of vevs.

\section{Correlators from gauge theory}

Since the scalars carry non-trivial $\mathcal{R}$-charge, the
$\mathcal{R}$-symmetry is in general broken on the Coulomb branch.
This is the well known phenomenon of spontaneous breaking of a global
symmetry in field theory,
and, therefore,
we expect massless poles in the $\mathcal{R}$-symmetry
current correlators for every broken symmetry generator,
which correspond to the massless Goldstone bosons.\footnote{See
also the pedagogical lectures on spontaneous symmetry breaking
\cite{beh} during this school.}
In order to investigate this issue we start
with the case of unbroken $\cal R$-symmetry where the vev's
corresponding to the six scalars of the theory are turned off.
The $\cal R$-symmetry currents
$J^a_\m$ are bilinear in the scalar fields
$X^i$, $i=1,2,\dots , 6$ and transform in the adjoint of $SO(6)$
\be
J^{kl}_\m = {1\ov g_{\rm YM}^2} T^{kl}_{ij} {\rm Tr}( X^i\del_\m X^j)\ + \
{\rm fermions}\ ,
\label{hl1}
\ee
where $T^{kl}_{ij}=\d_{ki}\d_{lj}-\d_{kj}\d_{ki}$
are the components of the $6\times 6$ matrices $T^{kl}$ of $SO(6)$.
The scalars $X^i$, being free fields, have the following two-point function
(in our conventions the field theory action has an overall
factor of $1/g_{YM}^2$)
\be
\langle X^i_{pq}(x) X^j_{rs}(0)\rangle = g_{\rm
YM}^2 \d^{ij} (\d_{qr} \d_{ps} -
{1\ov N} \d_{pq} \d_{rs}){1\ov r^2}\ , \qq p,q,r,s=1,2,\dots , N\ .
\label{twopoint}
\ee
After performing the Wick contractions we find the two-point
function of the currents
\be
\langle J_\m^{ij}(x) J_\n^{kl}(0)\rangle \sim N^2
(\d_{ik}\d_{jl}-\d_{il}\d_{jk})
(\square \delta_{\m\n} - \del_\m \del_\n) {1\ov r^4}\ ,
\label{cor1ne}
\ee
where we have kept only the leading term in the
$1/N$-expansion.\footnote{For finite $N$, the $N^2$ factor in
\eqn{cor1ne} is replaced by $N^2\!-\!1$
corresponding to the dimension of the $SU(N)$ group. We also note that
the contribution of the fermions only affects the result by an
overall $N$-independent numerical constant
which is not important for our purposes.}
This is indeed the correct result for the two point function which
also agrees with the AdS/CFT result \cite{fmmr}.

In the case that the symmetry is broken by turning on non-zero scalar
vev's, we replace the $X^i $ by $X^i_{\rm vev} + \delta X^i$,
where $X^i_{\rm vev}$ is defined in \eqn{vvee} and
the $\delta X^i$
have the same free field two-point function as in \eqn{twopoint}.
Besides the bilinear
term \eqn{hl1} the current contains now a term linear in fluctuating
fields
\be
\delta J^{kl}_\m = {1\ov g_{\rm YM}^2}
T^{kl}_{ij} {\rm Tr}( X_{\rm vev}^i\del_\m \delta X^j)\ ,
\label{hl11}
\ee
and the leading order correction to the conformal result \eqn{cor1ne} is
\be
\langle \delta J_\m^{ij} (x) \delta J_\n^{kl}(0)\rangle  \sim  {1\ov
g^2_{\rm YM}}
H^{ij,kl} \del_\m \del_\n {1\ov r^2}\ ,
\label{corrector}
\ee
where the group theoretical factor $H_{ij,kl}$ takes the form
\be
H^{ij,kl}  =  \d_{ik} A_{jl} - \d_{jk} A_{il} - \d_{il} A_{jk} +
\d_{jl} A_{ik} \ ,\qq  A_{ij} = \sum_{p=1}^N X^i_p X^j_p\ .
\label{ggrop}
\ee
It is clear that, in the ultraviolet where the vev's can be neglected,
the conformal
result \eqn{cor1ne} dominates, whereas in the infrared
the dominant term is
\eqn{corrector}.
The symmetric tensor $A_{ij}$ is given in terms of the
scalar vevs only and depends on their distribution. In the
following we think of the vevs $X^i_{\rm vev}$ as defining $N$ points in
$\mathbb{R}^6$.
In the large $N$ limit such a
discrete distribution can often be approximated by a
continuous one, as long as we work with energies
not too close to the vev values.
Furthermore, we will consider situations where the distribution spans only
a lower dimensional submanifold embedded in $\mathbb{R}^6$.
From (\ref{corrector}) we see that
the tensor $H^{ij,kl}$ contains all the important information about
the zero mass poles. It is antisymmetric in the indices $ij$ and $kl$
separately and symmetric under pairwise exchange.
Note that $A_{ij}$
is non-zero only
if both indices $i,j$ are along the vev-distribution.
That implies that $H^{ij,kl}=0$ if all indices correspond to directions
which are perpendicular to the distribution.

For the comparison with the dual supergravity
that we will perform later, it is
convenient to write down the general two point
function of the currents in a particular form.
In $x$-space it is given in terms of a function $G^{ij,kl}(x)$ as
\be
\langle J_\m^{ij}(x) J_\n^{kl}(0)\rangle = \frac{N^2}{32 \pi^4}
(\square \delta_{\m\n} - \del_\m \del_\n) \square G^{ij,kl}(x)\ ,
\label{cor1}
\ee
where the projector ensures transversality of the correlator.
In turn, the function $G^{ij,kl}(x)$ can be used to define a
function $H^{ij,kl}(k)$ in momentum space as
\be
G^{ij,kl}(x) = {1\ov 4\pi^2} \int d^4 k e^{i k\cdot x}
{H^{ij,kl}(k)\ov k^2} = {1\ov r}
\int^\infty_0 dk H^{ij,kl}(k) J_1(k r)\ ,
\label{cor2}
\ee
where the Bessel function $J_1(kr)$ is a result of the integration over
angular coordinates.
We cannot use $H^{ij,kl}(k)$ directly because
the correlator in $x$-space is too singular to be Fourier transformed
to momentum space.
However, by using differential regularization one can make sense of such
expressions by writing singular functions as derivatives of
less singular ones and then defining the Fourier transform by
formal partial integrations \cite{diffreg}.

We also note that the momentum space version of \eqn{corrector} can be
expressed in terms of a function $H^{ij,kl}(k)$ as
\be
H^{ij,kl}(k) \sim - \frac{1}{g^2_{\rm YM} N^2} \frac{H^{ij,kl}}{k^2} \ ,
\label{hhh}
\ee
where $H^{ij,kl}$ on the r.h.s. is defined in \eqn{ggrop}.
We emphasize that this is the interesting piece of the current correlator
that potentially gives rise to massless poles, i.e. Goldstone bosons,
depending on the details of $H^{ij,kl}$.

\subsubsection*{Some examples}

\no
\underline{The polygon}:
In this toy example we consider a discrete distribution of vevs in an
$N$-polygon whose vertices lie on a circle of radius
$r_0$ in the $1$-$2$ plane \eqn{lv}.
Computing the matrix elements $A_{ij}$ using the definition \eqn{ggrop}
is straightforward, and we find that the
only non-zero components are $A_{11} = A_{22} = N r_0^2/2$.
We note that in this case we obtain the same result even if we approximate the
discrete distribution by a continuous uniform distribution of vevs on the
circumference of the circle.

We now turn to the examples
with vev distributions on a disc and on a three-sphere, which
will be considered in section 3 (within a more general class of examples)
from the supergravity side
using the AdS/CFT correspondence.
In these cases a direct comparison with the free field calculation
can be performed and we will find precise agreement.

\medskip
\no
\underline{The three-sphere}: For a uniform distribution on a
three sphere of radius $r_0$
embedded in the 1-2-3-4 hyperplane
it is obvious that
$A_{ii}=N r_0^2/4$, for $i=1,2,3,4$ and zero otherwise. These
results are most easily derived
in the continuous approximation of the distributions.
Hence, using \eqn{ggrop}, \eqn{hhh} and the identities $g_{\rm YM}^2=g_s$ and
$R^4=4\pi g_s N$, we obtain
\be\label{sphere}
H_{\rm sphere}^\l(k) \sim - \frac{r_0^2}{R^4} \frac{\lambda}{k^2}\ ,
\ee
where the parameter $\lambda = 0, \frac{1}{2}$ and $1$ corresponds to
currents in the transverse directions (unbroken $SO(2)$), broken currents
in the coset and directions along the distribution (unbroken $SO(4)$),
respectively. 

\medskip
\no
\underline{The disc}:
For a uniform distribution on a disc in the 1-2 plane we have similarly that
$A_{ii}=N r_0^2/4$, for $i=1,2$ and zero otherwise.
Using \eqn{ggrop} and \eqn{hhh} we compute
\be\label{disc}
H_{\rm disc}^\l(k) \sim \frac{r_0^2}{R^4} \frac{\lambda - 1}{k^2}\ ,
\ee
where the parameter $\lambda = 0, \frac{1}{2}$ and $1$ corresponds to
currents along the distribution (unbroken $SO(2)$), broken currents
in the coset and directions orthogonal to the distribution (unbroken $SO(4)$),
respectively.


\section{The supergravity side}

On the dual supergravity side it is straightforward to write down the
relevant supergravity background for arbitrary points on the Coulomb branch.
The reason is that the vevs of the scalars are simply the positions of
the D3-branes in the transverse six-dimensional space \cite{kw}.
Furthermore, this
is a BPS configuration and the full solution is simply a superposition of
single D3-branes given by
\be\label{typeIIb}
ds^2 = \frac{1}{\sqrt{H}} dx_{||}^2 + \sqrt{H} \sum_{i=1}^6 dx_i^2 ~\ ,
\qq
H = 4 \pi g_s l_s^4 \sum_{p=1}^N \frac{1}{|\vec{x}-\vec{X}_p|^4}~,
\ee
where $x_{||}$ denotes flat worlvolume directions of the D3-brane.
While it is nice to have the most general solution, for practical
purposes of calculating e.g. current correlators, it is not very
useful and we will study a subspace of the Coulomb branch that (i) can
be studied using gauged supergravity, (ii) corresponds to continuous
distributions of D3-branes \cite{kraus,sfe1}
which means that the sum in \eqn{typeIIb} over
localized D3-branes has to be replaced by an integral over a specific
D3-brane density. This will allow us to make exact calculations in
several cases, which can be compared with the results discussed
at the end of section 2. In that respect, we mention the case
of the two center solution, where $N$ D3-branes are distributed evenly
in two stacks of branes. This case has been studied in \cite{MW}
and although it might look simpler than the ones we will encounter here,
it is actually difficult to perform any computation exactly.

The advantage of (i) is that we do not have to study the ten-dimensional
supergravity but we can restrict ourselves to a truncation of the full
theory that describes only the fields relevant for our problem.
In our case the relevant tool is five-dimensional
$\mathcal{N}=8$ gauged supergravity
\cite{5dgsu}
of which we actually only need a further truncation which includes
the metric, the $SO(6)$ gauge field and scalars in the
coset $SL(6,\mathbb{R})/SO(6)$.
The Lagrangian for these fields is
\be
\cL = \cL_{\rm scalar} + \cL_{\rm gauge}\ ,
\label{sugralagr}
\ee
where $\cL_{\rm scalar}$ denotes the pure gravity plus scalar sector
and $\cL_{\rm gauge}$ contains the kinetic term of the gauge fields
together with their interactions with gravity and the scalars.
The explicit form of the gravity plus scalar Lagrangian is
\be
\frac{1}{\sqrt{g}} \cL_{\rm scalar} = {1\ov 4} {\cal R} -
\frac{1}{16} {\rm Tr}
\left( \partial_{\widehat \m} M M^{-1} \partial^{\widehat \m}
M M^{-1} \right) - P
\ ,
\label{actionsc}
\ee
with the potential
\be
\label{potential}
P =  -{g^2\ov 32} \left[({\rm Tr} M)^2 - 2 {\rm Tr}(M^2)\right] \  ,
\ee
where $g$ is a mass scale, which is related to the $AdS_5$ radius
via $g=2/R$ with $R = (4 \pi g_s N)^{1/4} l_s$.
The scalar fields sit in a symmetric
traceless matrix $M^{ij}$ where the indices are in the fundamental
representation of $SO(6)$, i.e. these scalars transform in the
${\bf 20'}$. Supersymmetric domain-wall solutions of \eqn{actionsc}
preserving 16 supercharges together with four-dimensional
Lorentz invariance correspond to states on the Coulomb branch
of $\mathcal{N}=4$ SYM \cite{fgpw2}.
In this case the matrix $M$ can be
diagonalized using an $SO(6)$ gauge transformation and can be
parametrized by six scalar fields
\be\label{M}
M = {\rm diag} (e^{2 \beta_1}, \ldots, e^{2 \beta_6} )\ ,
\ee
obeying the constraint $\sum_{i=1}^6 \beta_i = 0$.
Alternatively one could use five independent scalar fields
$\a_I$, $I=1,2,\dots 5$ which are related to the $\b_i$'s by
$\b_i= \sum^5_{I=1} \l_{iI} \a_I$,
where $\l_{iI}$ is a $6\times 5$ matrix, with rows corresponding to the
fundamental representation of $SL(6,\IR)$.
The ansatz for the domain wall metric is
\be
ds^2 = e^{2 A(z)} (dz^2 + \eta_{\m\n} dx^\m dx^\n)
= dr^2 + e^{2A(r)} \eta_{\m\n} dx^\m dx^\n\ ,
\label{metriki}
\ee
where the relation between the coordinates $z$ and $r$ is such that
$dr=-e^A dz$.
The most general solutions is expressed in terms of an auxiliary
function $F(g^2 z)$, in terms of which the conformal factor and the
profiles of the scalars are \cite{bakas1}
\be
e^{2 A} =  g^2 (-F^\prime)^{2/3}\ ,\qq e^{2\b_i} = {f^{1/6}\ov F-b_i}\
,\qq f = \prod_{i=1}^6 (F-b_i) \ , \qq i=1,2,\dots, 6 \ .
\label{pro1}
\ee
The constants of integration are ordered as $b_1\geq b_2 \geq \dots \geq
b_6$ and the function $F$ obeys the differential equation
\be
(F^\prime)^4 = f\ .
\label{hd3}
\ee
Equating $n$ of the integration constants $b_i$ (or equivalently the
associated scalar fields $\b_i$) corresponds to preserving an $SO(n)$
subgroup of the original $SO(6)$ $\cR$-symmetry group.
In general the hypersurface $F=b_1$ corresponds to a
curvature singularity which, however, has the physical interpretation as
being the location of the distribution of D3-branes once we lift the
solution to a Type IIB background. Furthermore,
we can make contact with the disc and three-sphere distributions
discussed in section 2. The disc corresponds to setting the integration
constants $b_1=b_2=b_3=b_4$ and $b_5=b_6$ in which case the unbroken
$\cR$-symmetry group is $SO(2) \times SO(4)$ and the Goldstone bosons
corresponding to the broken symmetries reside in the coset
$SO(6)/(SO(2) \times SO(4))$.\footnote{See however the comments in the
paragraph starting after \eqn{add2}.}
For the three-sphere we have to choose
$b_1=b_2$ and $b_3=b_4=b_5=b_6$ while the unbroken symmetry group
and the coset are the same as for the disc. We note that the solutions
\eqn{metriki}, \eqn{pro1} can be lifted to type IIB solutions
of the form \eqn{typeIIb} and that, this class of solutions
does not include the uniform distribution of vevs on a circle.

Let us now add the gauge fields to the Lagrangian \eqn{actionsc}.
The partial derivatives in \eqn{actionsc} are replaced by
gauge-covariant ones
$\partial_{\widehat \m} M^{ij} \to
\partial_{\widehat \m} M^{ij} + g
(A_{\widehat \m}^{ik} M^{kj} +  A_{\widehat \m}^{jk} M^{ik})$, and
the gauge kinetic term has to be added
\be\label{gaugeaction}
\frac{1}{\sqrt{g}} \cL_{\rm gauge} =
-\frac{1}{8}
(M^{-1})^{ij}  (M^{-1})^{kl} F^{ik}_{\widehat \m \widehat \n}
F^{jl \widehat \m \widehat \n } \ ,
\ee
where $A_{\widehat \m}^{ij}$ and $F^{ij}_{\widehat \m \widehat \n}$
are anti-symmetric in $ij$.
Since we are interested in two-point functions we only keep terms
in \eqn{sugralagr} and \eqn{gaugeaction} which are
quadratic in the fluctuations $\delta A_{\widehat \m}^{ij}$ of the
gauge fields and the scalar fluctuations of the symmetric
unimodular matrix $\delta M^{ij}$.
Although we are only interested in the two point functions of the
gauge fields we have to keep scalar fluctuations since the gauge fields
couple to the off-diagonal scalar fluctuations $\delta M^{ij}$.
However, there are no couplings of the gauge fields to the metric
and the diagonal scalar fluctuations $\delta M^{ii}$
at quadratic order.
At this point it is crucial to distinguish between gauge fields
that correspond to unbroken symmetries for which $\beta_i = \beta_j$
and gauge fields corresponding to broken symmetries for which
$\beta_i \neq \beta_j$.

The case of unbroken symmetries is easier since
$\delta A_{\widehat \m}^{ij}$ fluctuations do not couple to
scalar fluctuations.
We can choose a convenient gauge $\delta A_z = 0$ and
the components along the world volume directions $A_\mu$ can be decomposed
into a transverse and a longitudinal polarization
$A_\mu = A_\mu^\bot + \partial_\mu \xi$.
The equation of motion for the physical modes $A_\mu^\bot$,
which obey $\partial^\mu A_\mu^\bot = 0$,
takes the form of a field equation
for a scalar field $\Phi$:
\be
\label{unbroken}
\del_z (e^B \del_z \Phi) - k^2 e^B  \Phi= 0 \ ,
\ee
with the definition
\be
B = A-2 (\b_i +\b_j)\ ,
\label{h93}
\ee
where we have omitted for notational convenience the $i,j$
dependence of $B$. Note that we have also performed a Fourier
transform in the $x^\mu$-directions with $k^2\equiv k_\mu k^\mu$.

For the broken symmetries $\beta_i \neq \beta_j$, however,
there are couplings between scalars and gauge fields and
things become more tricky.
It is a highly non-trivial fact that we can decouple the scalars via
the field redefinition \cite{jjcorr}
\be
A_{\widehat \m}^{ij} \to A_{\widehat \m}^{ij} + {1\ov g}\ \del_{\widehat \m}
\left(\d M_{ij}\ov e^{2\b_i} - e^{2\b_j}\right)\ ,\qq \b_i\neq \b_j\ ,
\label{frfe}
\ee
which has the form of an abelian gauge transformation.
Of course, this is nothing but the Higgs mechanism.
The Goldstone boson corresponding to the broken
gauge symmetries are {\it eaten} by the gauge bosons which
thereby obtain a mass.
Since the gauge fields are massive now we cannot eliminate degrees of
freedom by gauge fixing. In order to calculate the two-point
function we couple the gauge field to an external current via
$A^{ij}_{\widehat \mu} J^{\widehat \mu}_{ij}$ and require the
current to be covariantly conserved $D_{\widehat \mu} J^{\widehat \mu}_{ij}=0$.
The $A_z$ component decouples from the other modes
which in turn is related to the longitudinal component $\xi$.
Again, the decoupled equation for the transverse modes $A_\mu^\bot$,
takes the form of a massive scalar field equation
\be
\label{broken}
\del_z (e^B \del_z \Phi) -\left( k^2 e^B
+{1\ov 4} g^2 (b_i-b_j)^2 e^{-B} \right) \Phi= 0\ ,
\label{fiiin}
\ee
where the scalar $\Phi$ denotes any component
of $A^\bot_\m$.
For $\b_i=\b_j$, which implies $b_i = b_j$,
we recover from \eqn{broken} eq. \eqn{unbroken} that
describes the cases with unbroken symmetry.
Hence \eqn{broken} is the general equation that can be used
to calculate all current-current correlators.

Let us shortly pause here and admire the result. In the field theory
we study the breaking of a global symmetry, which is signalled by
the appearance of Goldstone bosons.
In the supergravity dual the global $\mathcal{R}$-symmetry
becomes a local gauge symmetry and we observe a different mechanism,
namely the Higgs phenomenon. The would be Goldstone boson becomes an
additional degree of freedom of the gauge field, i.e. the gauge
field becomes massive. Hence, we see that in the AdS/CFT correspondence
spontaneous symmetry breaking in the field theory corresponds to
Higgsing of a local gauge symmetry in the dual supergravity/string theory.
In the rest of these notes we complete the picture by showing
how the massless poles in the examples of section 2
can be reproduced
from a supergravity calculation of the current correlators
using \eqn{broken}.

We will follow the standard procedure of \cite{gkp,witten}
to determine the current-current correlators using \eqn{broken}.
We will work in Euclidean signature unless stated otherwise.
In order to proceed we need a complete set of eigenfunctions of
\eqn{broken}, which can be found explicitely only in a small
number of examples, namely for the disc and three-sphere distributions
\cite{fgpw2,brand1}.
Furthermore, we have to keep the solutions that blow up at the AdS
boundary since they correspond to operator
insertions \cite{gkp,witten}.
Finally, we have to evaluate
the on shell-value of the action ${1 \ov \kappa^2} \int d^5x {\cal L}$
with ${1 \ov \kappa^2} = \frac{N^2}{16 \pi^2}$ for solutions
$\Phi$ of \eqn{broken}. This yields the boundary term
\be
- \lim_{\epsilon \rightarrow 0} \frac{N^2}{32 \pi^2} e^B \Phi \del_z \Phi
\Big |_{z=\epsilon}^{z_{\rm max}} \equiv \frac{N^2}{16 \pi^2} k^2 H(k)   \ ,
\label{onshell}
\ee
where we have to normalize
$\Phi |_{z=\epsilon} = 1$ and take the limit
$\epsilon \to 0$ in \eqn{onshell}, which corresponds to the AdS boundary.
Re-introducing Lorentz and group theory indices properly,
we can present the current-current correlators in momentum space
schematically as
\be\label{momcorr}
\langle J_\m^{ij}(k) J_\n^{kl}(-k) \rangle = \frac{N^2}{8 \pi^2}
\left(\delta_{\m\n} - \frac{k_\m k_\n}{k^2}\right) k^4 \tilde{G}^{ij,kl}(k)\ ,
\ee
where a group theory factor and the momentum space
version of the projector, which guarantees that
the amplitude is transverse, have been included.
The expression \eqn{momcorr} is of course nothing but the
momentum-space analogue of \eqn{cor1}.

For the cases of D3-branes distributions over a disc and a three-sphere,
which were discussed in section 2, (\ref{broken}) can be solved
explicitly \cite{jjcorr}, but in general this is not possible.
But since we are mainly
interested in the Goldstone bosons we will take a shortcut
by solving (\ref{broken}) for small $k^2$ in order to
extract the massless poles.
It is useful to rewrite the equation \eqn{broken}
in terms of the variable $F$
\be
{d\ov dF} \left( (F-b_i)(F-b_j)  {d\Phi \ov dF}\right) -
k^2 { (F-b_i)(F-b_j)\ov f^{1/2}}\Phi - {b_{ij}^2\ov 4 (F-b_i)(F-b_j)}\Phi
=0 \ ,
\label{hjg}
\ee
where $F$ and $b_i$ were defined in
equation \eqn{pro1} and $b_{ij}=b_i - b_j$.
In order to extract the massless poles
it suffices to concentrate on the limit $k^2\to 0$,
where \eqn{hjg} can be solved exactly for any distribution.
This will give the leading contribution to the two-point function of
currents for large distances.
At the AdS boundary $F\to \infty$ we
impose the usual boundary condition
$\Phi\to 1$ corresponding to a point-like source.
Furthermore, we require $\Phi$ to be
smooth at the singularity $F=b_1$ in the interior.
In the following we use units where $g=2/R=1$.

\subsubsection*{Correlators and comparison with field theory}

In order to make a comparison between field theoretical and supergravity
results for the massless poles arising in the current correlators,
we need the group theoretical factor $H^{ij,kl}$ defined in \eqn{ggrop}.
It can be shown that \cite{jjcorr}
\be
A_{ij} = N b_{1j}\ \d_{ij}\ ,
\label{jefd}
\ee
where we defined $b_{ij}=b_i - b_j$. Consequently,
our distributions have a diagonal matrix $A_{ij}$.
Hence, the only non-zero independent components of the group
theoretical factor $H^{ij,kl}$ are $H^{ij,ij}$.
If all indices correspond to directions which are perpendicular to the
distribution then $H^{ij,kl}=0$, whereas if
all directions are along the distribution
$H^{ij,ij}=N (b_{1j} + b_{1i})$.
If we are in the coset one index is along the distribution (say $i$) and one
is orthogonal to it (say $j$), then one of the above terms is missing
and therefore $H^{ij,ij}=N b_{1i}$. This agrees perfectly with the
two special
cases of the disc and sphere distribution that we considered before.

\medskip
\no
\underline{Currents transverse to the distribution}:
In this case the indices of the current $i,j$  are such that
$b_i=b_j=b_1$. 
Demanding regularity at the singularity $F=b_1$
and imposing the normalization condition at the boundary gives
\be
\Phi=1\ .
\ee
Therefore \eqn{onshell} gives
\be H(k)=0\ .
\ee
As expected this agrees with the field theoretical results for
vev distributions on a three-sphere
(\ref{sphere}) and on a disc (\ref{disc}).

\medskip
\no
\underline{Currents longitudinal to the distribution}:
In this case the indices of the current are such that $b_i, b_j\neq b_1$.
Imposing regularity at the singularity at $F=b_1$
and the normalization condition at the boundary, we find
\be
\Phi = {1\ov b_{ij}} \left( b_{1j} \left({F-b_i\ov F-b_j}\right)^{1/2}
-b_{1i} \left({F-b_j\ov F-b_i}\right)^{1/2} \right)\ ,
\ee
from which, using \eqn{onshell}, we compute
\be
H(k) = - {b_{1i}+ b_{1j}\ov 4 k^2}\ .
\ee
A particularly interesting case is when $b_i=b_j\neq b_1$. Then the
above expressions reduce to
\be
\Phi = {F-b_1\ov F - b_i}\
\ee
and
\be
H(k) = - {b_{1i}\ov 2 k^2}\ .
\ee
The results for the sphere (\ref{sphere}) and disc (\ref{disc})
distributions correspond precisely to that result
with $b_{1i}=r_0^2/4$ ($b_1$ can be put to zero by a shift of the coordinate
$F$), for $i=1,2,3,4$ and $i=1,2$, respectively.

\medskip
\no
\underline{Currents in the coset}:
In this case the indices of the current are
$b_i=b_1$ and $b_j\neq b_1$.
Proceeding as before we find
\be
\Phi = \left({F-b_1\ov F-b_j}\right)^{1/2}
\ee
and
\be\label{llll}
H(k) = - {b_{1i}\ov 4 k^2}\ .
\ee
Setting $b_{1i}=r_0^2/4$, as above, one easily sees that the result
(\ref{llll}) agrees
with the field theory calculations for the disc (\ref{disc}) and
the three-sphere (\ref{sphere}), respectively.

\medskip
\no
\underline{Exact expressions for the sphere and disc distributions}:
In the case of a three-sphere distribution it is possible to compute exactly
the two-point function for current correlators in the supergravity side
\cite{jjcorr}, since \eqn{hjg} admits an appropriate solution in terms
of a hypergeometric function. Here we denote for completeness the expression
for
\ba
H^\l_{\rm sphere}(k) &= & {r_0^2\ov R^4} \frac{1-\l}{k^2}
\ +\ {1\ov 4}
 \Big(\psi\left((1+\D)/2\right) + \psi\left((1-\D)/2\right)+ 2 \g\Big)
\nonumber \\
& =& -{r_0^2\ov R^4} {\l\ov k^2}
\ +\ {1\ov 2} \sum_{n=1}^\infty {2 n + \tilde{k}^2 \ov n
(4n(n+1) + \tilde{k}^2)}  \  ,
\label{add1}
\ea
where $\tilde k^2\equiv k^2 R^4/r_0^2$ and $\psi(z)$
is the standard notation for
the derivative of the logarithm of the Gamma function $\G(z)$.
This expression has the correct behaviour
in the infrared for $k^2\to 0$ that we have already exhibited. In addition
it can be shown that it gives rise to the correct limit for the two-point
function in the ultraviolet.
The discrete spectrum of poles at $\tilde{k}^2 = - 4 n(n+1)$,
$n=1,2,\ldots$, corresponds precisely to the discrete mass eigenvalues for
the normalizable solutions of \eqn{hjg}.\footnote{We note that in general
the dilaton, transverse graviton and gauge field fluctuations have
degenerate spectra for our models \cite{brand2,jjcorr}.
This can be traced back to the
fact that the corresponding fields belong to the same $\cN=4$ supermutliplet.
For an explicit demonstration of this, in some related cases,
see \cite{massimo}.}

In the case of distribution of vevs on a disc we also obtain
\cite{jjcorr}
\ba
H^\l_{\rm disc}(k) & = &{r_0^2\ov R^4} {\l-1 \ov k^2}\
+ \ \ha \left( \psi\left((1+\D)/2\right)
+ \gamma\right)
\nonumber\\
& = & {r_0^2\ov R^4} {\l-1 \ov k^2}\ + \
\ha \int_0^\infty dt {e^{-t}-e^{-{\D+1\ov 2} t}\ov 1-e^{-t}}\ ,
\label{add2}
\ea
where $\D=\sqrt{\tilde k^2+1}$. This expressions also has the correct
behaviour in the infrared and ultraviolet limits. The branch cut for $\tilde
k^2=-1$ corresponds to a mass gap of the continuous
spectrum for the associated normalizable solutions (in the Dirac sense)
of \eqn{hjg}.

We close this section by explaining a subtlety of these results.
In the case of the three-sphere distribution we mentioned that the unbroken
$\mathcal{R}$ symmetry is $H = SO(2) \times SO(4)$, however we found
Goldstone bosons in the coset $SO(6)/H$, but also in the ``unbroken''
$SO(4)$ sector. The resolution of this discrepancy is that
the $SO(4)$ symmetry is accidental and is caused
by the approximation of a discrete
distribution by a smooth distribution over a three-sphere. Clearly,
the actual discrete distribution breaks this symmetry. Hence,
the Goldstone bosons live in the larger coset $SO(6)/SO(2)$.
Similar comments apply to the disk case where the role of $SO(2)$
and $SO(4)$ have to be interchanged. It is quite interesting that
the gravity dual seems to know about this and reproduces
the correct set of Goldstone bosons.

We have shown that the field theory and supergravity calculations
can be in excellent agreement even beyond the conformal limit.
The Goldstone bosons are
sensitive to the infrared physics of the field theory, which on the dual
supergravity side is captured by the interior of the geometry.
However, for states on the Coulomb branch the supergravity solutions
generically have naked singularities in the interior, hence
such good agreement is better than one could expect.
We have found that spontaneous
breaking of global symmetries of the field theory
translates on the supergravity side to the Higgs effect of local
gauge symmetries. It sounds somewhat counter intuitive, but the
massive bulk gauge fields corresponding to broken symmetries
conspire to reproduce the correct spectrum of Goldstone bosons
in the dual field theory.

\end{document}